\documentclass[11pt]{article}
\pdfoutput=1
\usepackage{fancybox,graphicx}
\usepackage{caption}
\usepackage{amsmath}
\usepackage[title]{appendix}
\def\sc {\scriptscriptstyle}
\def\spc {{\hskip 0.2mm}}
\def\x  {{\bf x}}
\def\c  {{\bf c}}
\def\C  {{\cal C}}
\def\f  {{\textsl f}}
\def\E  {{\mathcal E}}
\def\Z  {{\mathcal Z}}

\def\R  {{\mathcal R}}
\def\M  {{\text{M}}}
\begin{document}

\title{Sound Absorption and Dispersion in Dilute Polyatomic Gases:\break A Generalized Kinetic Approach}

\author{S. Fischer\footnote{Sebastian Fischer, B.Sc., Department of Physics, Technical University of Darmstadt, Germany}\, and W. Marques Jr.\\
Departamento de F\'{\i}sica, Universidade Federal do Paran\'a\\
Caixa Postal 19044, 81531-980, Curitiba, Brazil}

\date{}

\maketitle

\begin{abstract}

\noindent A generalized kinetic model equation which takes into account the frequency depence of the thermal conductivity is used to analyze 
the problem of sound propagation in dilute polyatomic gases. By comparing the theoretical results with some available experimental data 
we infer that our model equation provides a precise transition between low and high-frequency limits.

\end{abstract}

\begin{itemize}
\item[]{\bf Keywords:} \begin{minipage}[t]{0.7\textwidth}
polyatomic gases, kinetic model equation, frequency-dependent thermal conductivity, sound propagation
\end{minipage}
\end{itemize}

\section{Introduction}

\noindent Essentially, there are two different approaches to solve the problem concerning the propagation of plane harmonic waves in gases, 
namely: macroscopic and microscopic approaches. The macroscopic approach is based on conservation equations of mass, momentum and energy, and on 
the laws of Navier-Stokes and Fourier. Sound propagation predictions derived from this macroscopic approach are valid as long as the oscillation 
frequency of the acoustic wave is smaller than the mean molecular collision frequency \cite{Moraal}. In this case, sound dispersion is negligible, while the 
absorption per wavelength is proportional to the oscillation frequency and can be written as a sum of constributions due to viscous and thermal 
effects. Different generalizations of the usual macroscopic approach - trying to extend its validity to the high-frequency region - are reported 
in the literature \cite{Eu,Velasco,Zwanzig}. Among them, the most popular are those that take into account the frequency dependence of the transport coefficients. However, 
the sound propagation results derived from this approach give good agreement with the experimental data only for the phase velocity of the 
sound waves. When the oscillation frequency is comparable to the molecular collision frequency one has no recourse unless to turn to a microscopic 
approach, i.e., one has to use the Boltzmann equation. One of the earliest attempts to solve the problem of sound propagation 
in polyatomic gases based on a kinetic equation was made by Hanson, Morse and Sirovich \cite{Hanson}. By employing the method of Sirovich and Thurber for polyatomic 
kinetic models - which is based on a generalization of the Gross-Jackson procedure for monatomic gases - they derived dispersion relations for the 
four-moment and seven-moment approximations. Calculations performed by Hanson, Morse and Sirovich for nitrogen and oxygen show that their results are in 
some agreement with the experimental results of Greenspan \cite{Greenspan}.   

Our purpose in this paper is to study the propagation of sound waves in dilute polyatomic gases using a kinetic model equation which replaces 
the collision operator of the Boltzmann equation by a single relaxation time term \cite{Liu} and generalizes the classical hydrodynamic description by 
considering a frequency-dependent thermal conductivity. In particular, our expression for the generalized thermal conductivity follows from 
the Maxwell-Cattaneo law of heat conduction which can be written as a constitutive equation for the heat flux vector with an exponential memory 
kernel. By applying the normal mode method to our generalized model equation, it is possible to derive a dispersion relation which can be used to 
determine the phase velocity and the attenuation coefficient of the acosutic mode. Comparison of the theoretical results with the acoustic measurements 
performed by Greenspan in oxygen and nitrogen shows that our generalized kinetic model equation for dilute polyatomic gases provides a precise transition 
between low and high-frequency limits.

We organize the paper as follows: in Section 2 we derive a generalized kinetic model equation for dilute polyatomic gases which takes into account the 
frequency dependence of the thermal conductivity. The sound wave propagation problem is analyzed in Section 3, while in Section 4 we compare 
the theoretical results with some available experimental data. Finally, in Section 5, we finish with some concluding remarks. 

Cartesian notation for tensors with the usual summation convention is used. Furthermore, angular parentheses around indices denote traceless symmetrization.

\section{Generalized Model Equation}

\noindent In the classical kinetic theory developed by Taxman \cite{Taxman} for dilute polyatomic gases, the one-particle distribution function $f(\x,\c,t,s)$ is defined 
in such a way that $f(\x,\c,t,s)\spc d\c\spc ds$ gives the number of molecules at position $\x$ and time $t$ with molecular velocities between $\c$ and 
$\c+d\c$ and with internal degrees of freedom between $s$ and $s+ds$. By neglecting external forces, the one-particle distribution function satisfies 
the Boltzmann equation \cite{Chapman}
\begin{gather}
\frac{\partial f}{\partial t}+c_i\frac{\partial f}{\partial x_i}={\cal C}(f,f),
\end{gather}
where $\C (f,f)$ is the Boltzmann collision operator. The collision operator describes the rate of change of the distribution function 
due to molecular collisions and obeys the following collisional invariant conditions \cite{Kremer}:
\begin{gather}
\int m\spc\spc {\cal C}(f,f)\spc d\c\spc ds=0,
\end{gather}
\begin{gather}
\int mc_i\spc {\cal C}(f,f)\spc d\c\spc ds=0,
\end{gather}
\begin{gather}
\int \left(\frac{mc^2}{2}+\E_s\right) {\cal C}(f,f)\spc d\c\spc ds=0,
\end{gather}
where $m$ is the molecular mass and $\E_s$ denotes the energy associated with the internal degrees of freedom. Conditions (2)-(4) correspond to the 
conservation of mass, momentum and energy during collisions, respectively.

The mathematical complexity of the collision operator $\C (f,f)$ to analyze time-dependent problems like sound propagation and light-scattering in 
dilute polyatomic gases is usualy avoided by replacing it by a single relaxation-time term of the form 
\begin{gather}
{\cal C}(f,f)=-\frac{\left(f-f_r\right)}{\tau},
\end{gather}
where $\tau$ is an effective relaxation time and $f_r$ is a reference distribution function which satisfies the main physical properties 
of the Boltzmann collision operator. Recently, an expression for the reference distribution function was obtained by Marques Jr. \cite{Marques} by requiring the 
Chapman-Enskog solution of the kinetic model equation to be consistent with the classical Navier-Stokes-Fourier description. 

In the usual Navier-Stokes-Fourier theory 
a macroscopic state of a dilute polyatomic gas is characterized by the fields of mass density 
\begin{gather}
\rho=\int m f d\c\spc ds,
\end{gather}
flow velocity 
\begin{gather}
v_i=\rho^{-1}\int m c_i f d\c\spc ds,
\end{gather}
and temperature 
\begin{gather}
T=(\rho c_v)^{-1}\int \left(\frac{mC^2}{2}+\E_s\right) f d\c\spc ds ,
\end{gather}
while the pressure tensor
\begin{gather}
p_{ij}=\int m C_iC_j f d\c\spc ds
\end{gather}
and the heat flux vector
\begin{gather}
q_i=\int \left(\frac{mC^2}{2}+\E_s\right)C_i f d\c\spc ds
\end{gather}
are given, respectively, by the following constitutive relations \cite{Ferziger} 
\begin{gather}
p_{ij}=\left(p-\eta_{\sc 0}\spc \frac{\partial v_r}{\partial x_r}\right)\delta_{ij}
-2\spc \mu_{\sc 0}\spc \frac{\partial v_{\langle i}}{\partial x_{j\rangle}}
\end{gather}
and
\begin{gather}
q_i=-\lambda_{\sc 0}\spc \frac{\partial T}{\partial x_i},
\end{gather}
where  $c_v$ is the total specific heat at constant volume, $C_i=c_i-v_i$ is the peculiar velocity, $p$ is the gas pressure, 
$\mu_{\sc 0}$ is the shear viscosity, $\eta_{\sc 0}$ is the bulk viscosity and $\lambda_{\sc 0}$ is the thermal 
conductivity. By taking the effective relaxation time $\tau$ equal to the stress relation time $\tau_{\sc s}=\mu_{\sc 0}/p$,
the reference distribution function for dilute polyatomic gases which is compatible with the classical 
Navier-Stokes-Fourier theory reads
\begin{gather}
f_r=f^{\sc (0)}\left\{1+\tau\left(1-\frac{\f_{\spc \sc 0}}{\gamma}\right)\left[\left(\frac{mC^2}{2kT}-\frac{5}{2}\right)
+\left(\frac{\E_s}{kT}-\frac{\E}{kT}\right)\right]\frac{C_i}{T}\frac{\partial T}{\partial x_i}\right.\nonumber\\[4mm]
\left.+\,\tau\left(1-Z_{\sc 0}\right)\left[\left(\frac{5}{3}-\gamma\right)\left(\frac{mC^2}{2kT}-\frac{3}{2}\right)-\left(\gamma-1\right)
\left(\frac{\E_s}{kT}-\frac{\E}{kT}\right)\right]\frac{\partial v_r}{\partial x_r}\right\},
\end{gather}
where
\begin{gather}
f^{\sc (0)}=\frac{\rho}{m}\left(\frac{m}{2\pi k T}\right)^{\!\! 3/2}\exp\left(-\frac{ mC^2}{2 kT}\right)
\frac{\displaystyle \exp\left(-\E_s/kT\right)}{\Z}
\end{gather}
is the local equilibrium distribution function and
\begin{gather}
\Z=\int \exp\left(-\E_s/kT\right)ds
\end{gather}
is the partition function associated with the internal degrees of freedom. Moreover, $\gamma$ is the specific heat ratio, 
$\f_{\spc\sc 0}=\lambda_{\sc 0}/\mu_{\sc 0} c_v$ is the Eucken factor, $Z_{\sc 0}=(\eta_{\sc 0}/\mu_{\sc 0})/(5/3-\gamma)$ is the internal relaxation number 
(i.e., the mean number of molecular collisions required for the translational and internal degrees of freedom to come to thermal equilibrium \cite{Boley})
and 
\begin{gather}
\E=\frac{1}{\Z}\int \E_s\spc \exp\left(-\E_s/kT\right)ds
\end{gather}
is the mean internal energy. 

By analyzing time-dependent problems like sound propagation and light scattering in dilute polyatomic gases, Marques Jr. \cite{Marques} 
was able to determine the range of applicability of his kinetic model equation. Comparison of theoretical results with 
available experimental data in nitrogen, oxygen, carbon dioxide and methane shows that the reference distribution function (13) is 
valid as long as the external oscillation frequency is smaller than the relaxation frequency for internal and translational degrees 
of freedom to come to thermal equilibrium.

It is possible to extend the validity of the kinetic model equation to the high-frequency region by considering a generalization of 
the classical hydrodynamic description. Certainly, the most popular approach that generalizes the hydrodynamic behaviour of simple fluids
is based on the Maxwell-Cattaneo law of heat conduction \cite{Velasco}
\begin{gather}
\tau_{\sc q}\frac{\partial q_i}{\partial t}+q_i=-\lambda_{\sc 0}\frac{\partial T}{\partial x_i},
\end{gather}
where the relaxation time $\tau_{\sc q}$ gives us a measure of the time interval spent by the heat flux vector to achieve a stationary value. 
It is well known \cite{Vernotte,Velasco2} that the Maxwell-Cattaneo equation (17) can be written as 
\begin{gather}
q_i=-\int_0^t\frac{\lambda_{\sc 0}}{\tau_{\sc q}}
\,\exp\left(-\frac{t-t^\prime}{\tau_{\sc q}}\right)\frac{\partial T(\x,t^\prime)}
{\partial x_i}\,dt^\prime,
\end{gather}
i.e., as a constitutive equation with an exponential memory kernel in such a way that a generalized thermal conductivity can be 
defined as  
\begin{gather}
\lambda (t-t^\prime)=\frac{\lambda_{\sc 0}}{\tau_{\sc q}}
\,\exp\left(-\frac{t-t^\prime}{\tau_{\sc q}}\right).
\end{gather}
Since sound wave and light scattering solutions are found by Fourier transformation 
in space and time, we verify from (19) that this generalization introduces a frequency-dependent thermal conductivity
\begin{gather}
\lambda (\omega)=\frac{\lambda_{\sc 0}}{1+i\,\omega\,\tau_{\sc q}}.
\end{gather}
Furthermore, based on the work of Zwanzig \cite{Zwanzig}, it is possible to modify the constitutive relation (11) for the pressure tensor 
to take into account the frequency dependence of the bulk viscosity. Aiming to calculate the spectral distibution of scattered light in a 
one-component fluid whose molecules have internal degrees of freedom weakly coupled to their translational degrees of freedom by a single relaxation 
time process, Mountain wrote the pressure tensor as \cite{Mountain2}
\begin{gather}
p_{ij}=\left(p-\eta_{\sc 0}\spc \frac{\partial v_r}{\partial x_r}
-\int_0^t\frac{\eta_{\sc 0}}{\tau_{\sc \ast}}\spc \exp\left(-\frac{t-t^\prime}{\tau_{\sc \ast}}\right)\frac{\partial v_r(\x,t^\prime)}{\partial x_r}\spc dt^\prime
\right)\delta_{ij}-2\spc \mu_{\sc 0}\spc \frac{\partial v_{\langle i}}{\partial x_{j\rangle}},
\end{gather}
where $\tau_{\sc \ast}$ is the relaxation time. Expression (21) shows that in the so-called weak coupling limit the bulk viscosity consists of two parts, a frequency 
independent one due to translational motions and a frequency dependent one which is related to the exponential decay of the energy in internal modes as a result of interactions 
with translational modes. Comparison between theory and experiments shows that the generalized hydrodynamical model proposed by Mountain does not always 
apply since in some fluids the relaxation of the internal degrees of freedom involves more than one relaxation time. 

Based on these facts we construct in this work a generalized kinetic model equation for dilute polyatomic gases which takes into account the frequency 
dependence of the thermal conductivity via the Maxwell-Cattaneo law, but disregards the frequency dependent part of the bulk viscosity. In addition, we 
can neglect the frequency dependence of the shear viscosity since (as pointed out by Zwanzig \cite{Zwanzig}) the shear tensor is related to the transport 
of momentum, not to the transport of energy and, therefore, the internal degrees of freedom of the molecules do not contribute to the shear viscosity. 
For the construction of our generalized kinetic model equation we start by assuming that the reference distribution function is 
given by the expression 
\begin{gather}
f_r=f^{ \sc (0)}\biggl\{1+A\biggl(\frac{\E_s}{kT}-\frac{\E}{kT}\biggr)\!+A_iC_i+A_{rr}\biggl(\frac{mC^2}{2kT}-\frac{3}{2}\biggr)\!
+A_{\langle ij\rangle}C_iC_j+A_{irr}\biggl(\frac{mC^2}{2kT}+\frac{\E_s}{kT}\biggr)C_i\biggr\},
\end{gather}
where $A$, $A_i$, $A_{rr}$, $A_{\langle ij\rangle}$ and $A_{irr}$ are expansion coefficients that depend on position and time through the basic fields. 
As described in details by Marques Jr. \cite{Marques}, the application of the Chapman-Enskog method \cite{Chapman} to solve the kinetic model equation  
allows the determination of the expansion coefficients appearing in the above reference distribution function. Hence, by considering 
the constitutive relations (11) and (18) we get
\begin{gather}
\left(\frac{5}{3}-\gamma\right)\!A+\left(\gamma-1\right)\!A_{rr}=0,\nonumber\\[3mm]
A_i+\left(\frac{5}{2}+\frac{\E}{kT}\right)\!A_{irr}=0,\nonumber\\[3mm]
A_{\langle ij\rangle}=0,\\[3mm]
A_{rr}=\frac{\mu_{\sc 0}}{p}\left(\frac{5}{3}-\gamma\right)(1-Z_{\sc 0})\spc \frac{\partial v_r}{\partial x_r},\nonumber\\[3mm]
A_{irr}=\frac{\mu_{\sc 0}}{p\spc T}\left[\frac{\partial T}{\partial x_i}-\frac{1}{\tau_{\sc q}}\int_0^t\frac{\f_{\spc \sc 0}}{\gamma}
\,\exp\left(-\frac{t-t^\prime}{\tau_{\sc q}}\right)\frac{\partial T(\x,t^\prime)}
{\partial x_i}\,dt^\prime\right].\nonumber
\end{gather}
Finally, by combining expressions (1), (5), (22) and (23) we can write our generalized kinetic model equation for dilute polyatomic gases as 
\begin{gather}
\frac{\partial f}{\partial t}+c_i\frac{\partial f}{\partial x_i}=-\frac{p}{\mu_{\sc 0}}\left(f-f^{\sc (0)}\right)
+f^{\sc (0)}(1-Z_{\sc 0})\left[\left(\frac{5}{3}-\gamma\right)\left(\frac{mC^2}{2kT}-\frac{3}{2}\right)-(\gamma-1)
\left(\frac{\E_s}{kT}-\frac{\E}{kT}\right)\right]\frac{\partial v_r}{\partial x_r}\nonumber\\[4mm]
+f^{\sc (0)}\left[\left(\frac{mC^2}{2kT}-\frac{5}{2}\right)+\left(\frac{\E_s}{kT}-\frac{\E}{kT}\right)\right]\frac{C_i}{T}
\left[\frac{\partial T}{\partial x_i}-\int_0^t\frac{\f\spc\spc (t-t^\prime)}{\gamma}\,\frac{\partial T(\x,t^\prime)}
{\partial x_i}\,dt^\prime\right],
\end{gather}
where
\begin{gather}
\f\spc\spc (t-t^\prime)=\frac{\f_{\spc\sc 0}}{\tau_{\sc q}}\spc \exp\left(-\frac{t-t^\prime}{\tau_{\sc q}}\right)
\end{gather}
is the generalized Eucken factor. Being an approximation to the Boltzmann equation for dilute polyatomic gases, equation (24) must fulfill 
two basic properties of the true collision term of the Boltzmann equation, namely: (i) for all sumation invariants the collision term (5) must satisfies 
the conditions (2)-(4) and (ii) the tendency of the distribution function to equilibrium (or equivalently the H-theorem) must hold, i.e., $\int \ln f\spc 
\C(f,f)\spc d{\bf c}\spc ds \leq 0$. Because the construction of our model equation is based on constrains (2)-(4), the first property is verified. Concerning 
the H-theorem, we show in Appendix A that our model equation also satisfies this requirement.

Closing this section, we call attention to the fact that the application of the above generalized kinetic model equation to study time-dependent 
problems in dilute polyatomic gases only requires the specification of the ratio of the specific heats $\gamma$, the Eucken factor $\f_{\spc\sc 0}$ 
and the internal relaxation number $Z_{\sc 0}$.

\section{Absorption and Dispersion of Sound}

\noindent As an application of the generalized kinetic model equation derived in the previous section, we study the problem concerning the 
propagation of a plane harmonic wave in dilute polyatomic gases. To simplify this study we consider the sound propagation process 
in the linear regime near equilibrium and assume that the sound wave moves along the $x$-axis. Hence, we write the distribution function 
as
\begin{gather}
f=f_{\sc 0}\left\{1+\psi\spc \exp\left[\spc i\spc (\kappa x-\omega t)\right]\right\},
\end{gather}
where $f_{\sc 0}$ is the absolute equilibrium distribution function, $\kappa=\omega/v+i\spc \alpha$ is the complex wavenumber, 
$\omega$ is angular oscillation frequency, $v$ is the phase velocity, $\alpha$ is the attenuation coefficient
and $\psi=\psi (\c)$ is a function that depends only on the molecular velocity. In sound propagation problems 
the complex wavenumber $\kappa$ is determined as a function of the oscillation frequency $\omega$ via 
solution of the dispersion relation (for details see \cite{Marques, Marques3, Marques2})
\begin{gather}
\text{det}\left(\R {\bf M}^{\sc (0)}-i\left(\kappa v_{\sc 0}/\omega\right)\left({\bf I}-{\bf M}^{\sc (1)}\right)\right)=0,
\end{gather}
where $\R=p_{\sc 0}/\mu_{\sc 0}\omega$ is the rarefaction parameter, $v_{\sc 0}=\sqrt{2kT_{\sc 0}/m}$ is the equilibrium thermal velocity 
and ${\bf I}$ is the identity matrix. Moreover, ${\bf M}^{\sc (0)}$ and ${\bf M}^{\sc (1)}$ are $3\times 3$ matrices whose elements - given in Appendix B - depend on the dimensioless parameter $z=(\omega/\kappa v_{\sc 0})(1+i\spc \R)$ and on the plasma 
dispersion function \cite{Fried2}
\begin{gather}
W(z)=\frac{1}{\sqrt{\pi}}\int_{-\infty}^{+\infty}\frac{\exp\left(-t^2\right)}{t-z}\spc dt.
\end{gather}  

The acoustic solution of the dispersion relation (27) in the low-frequency limit can be determined by expanding the dimensionless wave number 
$\kappa v_{\sc 0}/\omega$ in powers of $1/\R$. By retaining terms up to first order we obtain
\begin{equation}
\begin{split}
\frac{\kappa c_{\sc 0}}{\omega} & = 1 + \dfrac{i}{2\spc \gamma\spc \R} \left[ \dfrac{4}{3} + \left( \dfrac{5}{3} - \gamma \right) Z_{\sc 0} + \dfrac{\gamma - 1}{\gamma} \spc \f_{\spc \sc 0} \right]\\[0.5cm]
& = 1 + i \dfrac{\omega}{2\spc \gamma \spc p_{\sc 0}}\left(  \frac{4}{3} \mu_{\sc 0} + \eta_{\sc 0} + \frac{\gamma - 1}{\gamma}\frac{\lambda_{\sc 0}}{c_{v}}  \right).
\end{split}
\end{equation}
in the low-frequency limit. Expression (29) shows us that the dispersion of sound is negligible 
in the low-frequency region, while the sound absorption per wavelength is proportional to the sound oscillation frequency and can be written as a sum 
of viscous and thermal effects. Note that an additional contribution coming from internal relaxation appears in the absorption coefficient of polyatomic 
gases, a fact that allows us to determine the bulk viscosity by comparing the theoretical results with available experimental data. Moreover, it is 
important to mention that the classical Navier-Stokes-Fourier description leads to the same expressions for sound absorption and dispersion 
in the low-frequency limit. This result is in complete agreement with our expectations, since we have required that the Chapman-Enskog solution 
of our kinetic model equation to be consistent with the usual macroscopic approach. 

\newpage

\section{Comparison with Experiments}

\begin{figure}[h!]
\centering
\includegraphics[width=11.0cm]{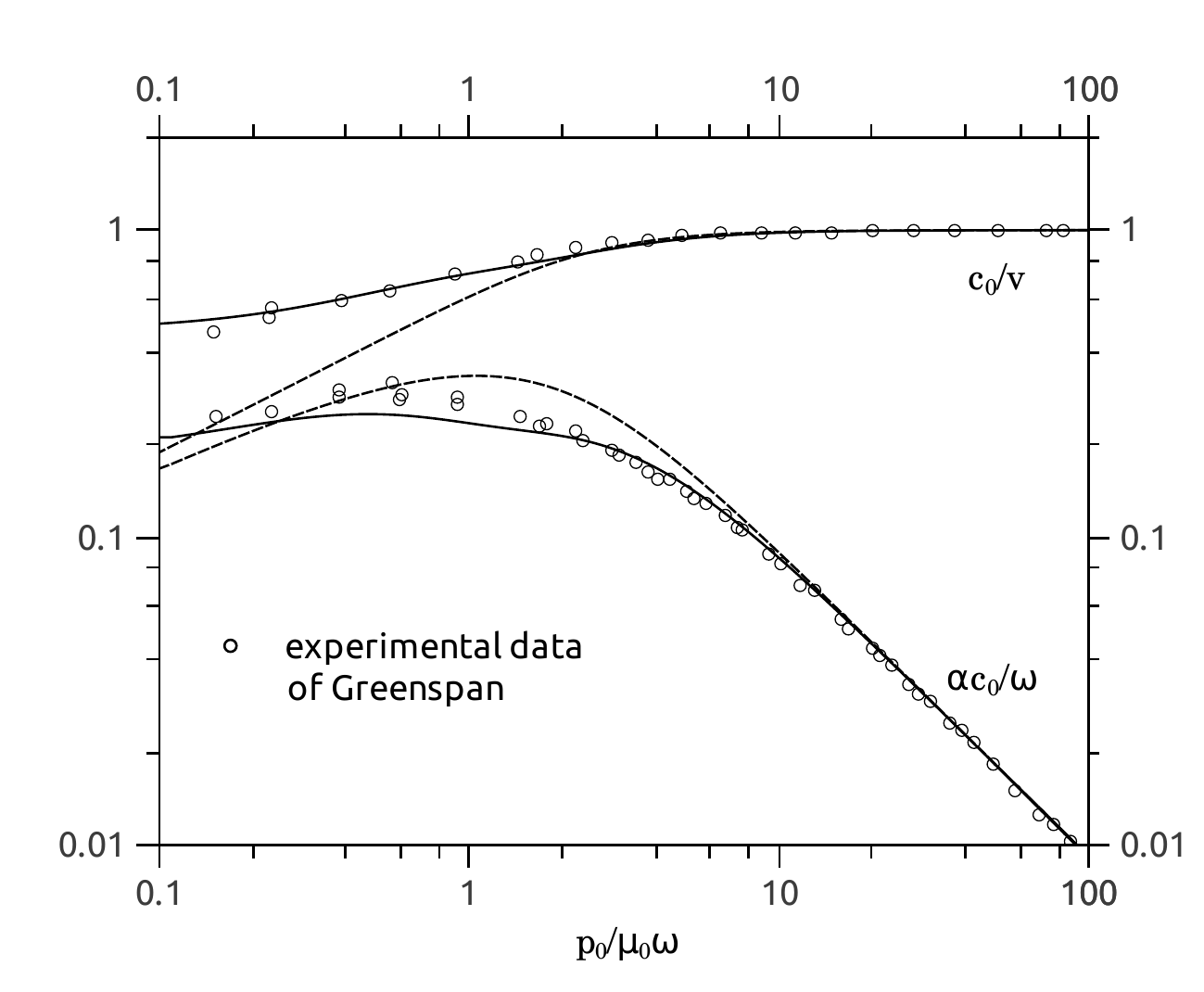}
\caption{Attenuation factor $\alpha c_{\sc 0}/\omega$ and reciprocal speed ratio $c_{\sc 0}/v$ as a function of the rarefaction parameter $\mathcal{R}=p_{\sc 0}/\mu_{\sc 0}\omega$ for nitrogen at 300K. The predictions of our generalized kinetic model equation~(\textbf{---}) are compared with the experimental data of Greenspan and the results of the usual macroscopic approach (-\,-\,-).}
\label{Fig3}
\end{figure}

\begin{figure}[h!]
\centering
\includegraphics[width=11.0cm]{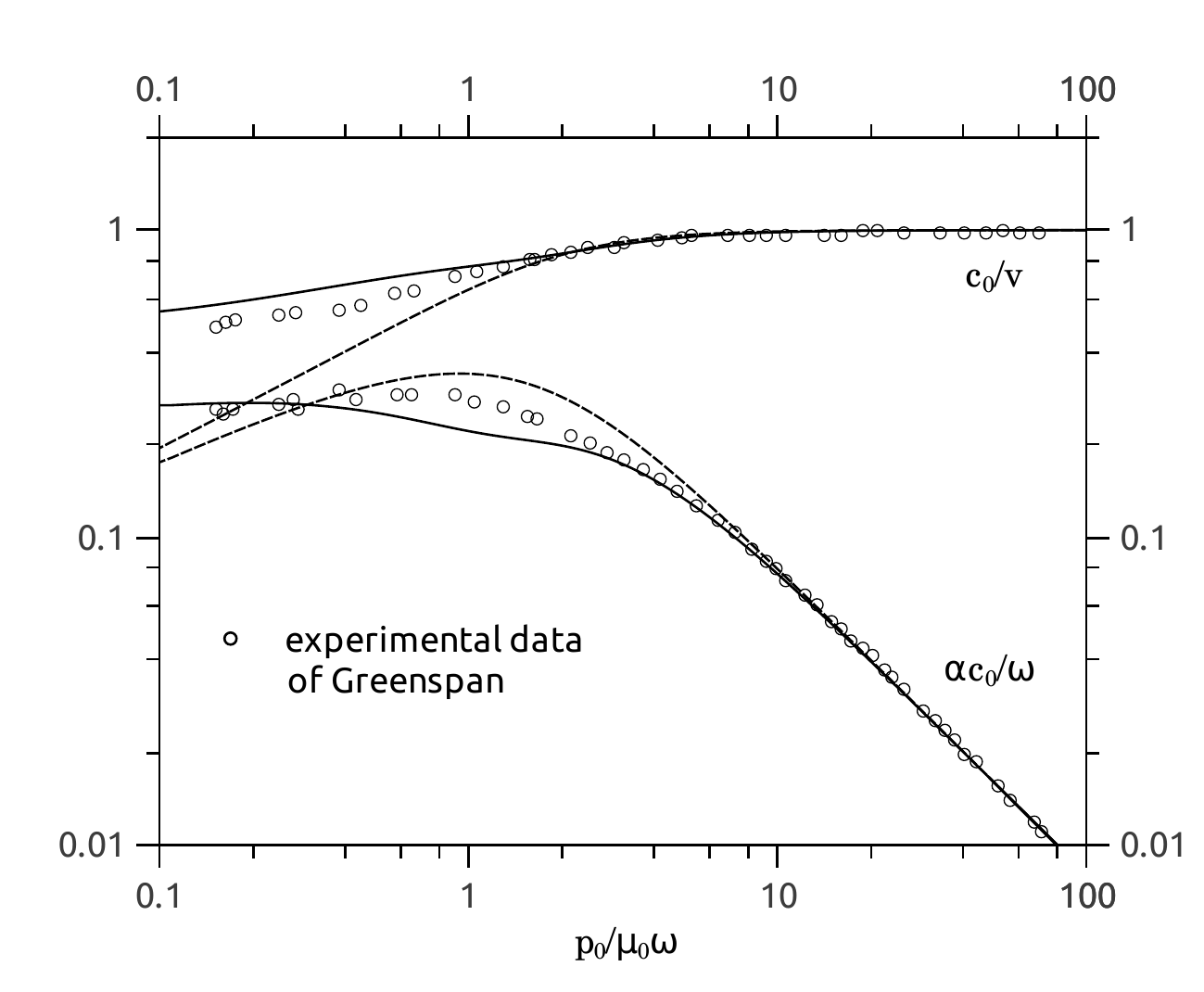}
\caption{Attenuation factor $\alpha c_{\sc 0}/\omega$ and reciprocal speed ratio $c_{\sc 0}/v$ as a function of the rarefaction parameter $\mathcal{R}=p_{\sc 0}/\mu_{\sc 0}\omega$ for oxygen at 300~K. The predictions of our generalized kinetic model equation~(\textbf{---}) are compared with the experimental data of Greenspan and the results of the usual macroscopic approach (-\,-\,-).}
\label{Fig4}
\end{figure}

In order to test the validation and the range of applicability of the generalized kinetic model equation proposed in this paper, we compare in 
Fig. \ref{Fig3} and Fig. \ref{Fig4} the sound wave absorption factor $\alpha c_{\sc 0}/\omega$ and the reciprocal speed ratio $c_{\sc 0}/v$ calculated 
from the solution of the corresponding dispersion relation~(27) with the acoustic measurements of Greenspan in nitrogen and oxygen. These sound 
propagation measurements were made at a temperature of 300~K in a 11~MHz double-crystal interferometer for different values of the gas pressure. 
The sound wave absorption factor was obtained by a determination of the logarithmic decrement in the sound level of the signal as a function of 
the traveled sound path, whereas the reciprocal speed ratio was determined by measuring the phase difference between a direct signal from the 
driving oscillator and the signal received at the receiver ditto as a function of the sound path \cite{Greenspan,Kahn}. For the numerical calculations 
we used the following material parameters:

\vskip 0.75truecm

\centerline{
\begin{tabular}{ccccccc}
\hline
\hline
Gas &  & $\gamma$ &  & $\f_{\spc \sc 0}$ &  & $Z_{\sc 0}$ \\
\hline
$\text{N}_2$ &  & 1.40 &  & 1.97 &  & 2.5\\
$\text{O}_2$ &  & 1.39 &  & 1.95 &  & 1.3\\
\hline
\hline
\end{tabular}}

\vskip 0.75truecm

\noindent The values of the specific heat ratio $\gamma$ and the Eucken factor $\f_{\spc \sc 0}$ were obtained by using the values 
of $c_{\sc v}$, $c_{\sc p}$, $\mu_{\sc 0}$ and $\lambda_{\sc 0}$ given in the CRC Handbook of Chemistry and Physics \cite{CRC}, while the values 
of the internal relaxation number $Z_{\sc 0}$ were obtained by using expression (29) 
to fit the sound absorption experiments of Greenspan in the low-frequency limit.

In Fig. \ref{Fig3} and Fig. \ref{Fig4} the attenuation factor $\alpha c_{\sc 0}/\omega$ and the reciprocal speed ratio $c_{\sc 0}/v$ are shown on a double logarithmic scale as a function of the rarefaction parameter $\mathcal{R}=p_{\sc 0}/\mu_{\sc 0}\omega$ for nitrogen and oxygen, respectively. The solid line represents the theoretical sound propagation results derived from our generalized kinetic model equation, while the dashed line represents the theoretical results derived from the usual macroscopic (hydrodynamic) approach based on the laws of Navier-Stokes and Fourier with frequency independent transport coefficients. The open circles are the experimental data of Greenspan for sound wave absorption (lower curve) and dispersion (upper curve).

We verify from Figs. \ref{Fig3} and \ref{Fig4} that in the low-frequeny region ($\mathcal{R}\gg 1$) the theoretical sound propagation results derived from our generalized kinetic model equation are in complete agreement with the experimental data. Moreover, the predictions for sound wave absorption and dispersion are exactly the same as obtained by the macroscopic approach. Since we required the Chapman-Enskog solution of our model equation to be consistent with the usual macroscopic description, this result is in complete agreement with our expectations. In the transition region ($1<\mathcal{R}<10$) we note that the kinetic theory proposed in this paper yields better results than the usual macroscopic approach. Whereas the latter predicts a value for the attenuation factor which is about 23\% greater than the experimental value, the predictions derived from our generalized kinetic model equation are in good agreement with the absorption data up to a value of the rarefaction parameter of approximately~2. For smaller values of the rarefaction parameter, i.e., in the high-frequency region we observe that the usual macroscopic approach yields a qualitatively correct result for the absorption curve but fails to describe the dispersion data. In contrast, the theoretical results obtained from our generalized kinetic model equation for small $\mathcal{R}$ are in good agreement with both, especially with the sound wave dispersion curve of nitrogen.
Further, we can conclude from Figs. 1 and 2 that the agreement between theory and experiment in the high-frequency limit is better for nitrogen than for oxygen, i.e., for gases with higher values of $Z_{\sc 0}$ which can be explained in terms of the internal relaxation time $\tau=\mu_{\sc 0} Z_{\sc 0}/p$. A higher value of $Z_{\sc 0}$ means that the thermal relaxation process between translational and internal degrees of freedom runs more slowly and, therefore, has a greater influence on the propagating sound wave which is appropriately described by our generalized kinetic model equation.

The comparison presented in Figs. \ref{Fig3} and \ref{Fig4} indicates that there are still opportunities for further enhancements in the kinetic theory, such as including the complete frequency dependence of all transport coefficients or considering non-linearities in the model equation, but all in all, we want to remark that the kinetic theory is well suited to describe sound propagation in dilute polyatomic gases over a wide range of frequencies and provides a precise transition between low and high-frequency limits. Whereas the macroscopic approach becomes unable to describe the acoustic measurements in the high-frequency region, the kinetic model equation derived in this work still provides a qualitatively correct description of the experimental data, even for values of the rarefaction parameter smaller than the unity. 

\section{Conclusion and Outlook} 
In the present work, we are concerned with the time-dependent problem of sound propagation in dilute polyatomic gases with classical internal degrees of freedom. Since the macroscopic (hydrodynamic) approach fails to describe the experimental data in the high-frequency region one needs to turn to a microscopic (kinetic) approach based on the Boltzmann equation. Assuming that the Boltzmann collision operator can be replaced by a single relaxation-time term in order to eliminate its mathematical complexity, we are able to derive a generalized kinetic model equation for dilute polyatomic gases which explicitly takes into account the frequency dependence of the thermal conductivity. By comparing the theoretical sound propagation results derived from our generalized kinetic model equation with the experimental data, we observe that the kinetic theory is well suited to describe sound propagation in dilute polyatomic gases over a wide range of frequencies. Whereas the macroscopic approach becomes unable to describe the acoustic measurements in the high-frequency region, the generalized kinetic model equation proposed in this work still provides a qualitatively correct description of the experimental data, even for values of the rarefaction parameter smaller than the unity. The presented comparison also indicates that there are still opportunities for further enhancements in the kinetic theory. One of the next steps for future work on this topic would be to include the complete frequency dependence of all transport coefficients into the kinetic model equation. 

Finally, we want to close this section by remarking that for the application of our generalized kinetic model equation to a specific sound propagation problem no further information about the potential energy of interaction between the gas molecules is required because the acoustic properties are completely characterized by three parameters, namely the ratio of the specific heats $\gamma$, the Eucken factor $\f_{\sc 0}$ and the internal relaxation number $Z_{\sc 0}$. Values of these parameters can easily be found in the literature or fitted to experimental data in the low-frequency limit.

\newpage
\begin{appendices}
\renewcommand{\theequation}{\thesection.\arabic{equation}}
\setcounter{equation}{0}

\section{}
\noindent In order to prove the $H$-theorem, let us rewrite our generalized kinetic model equation for dilute polyatomic gases as
\begin{gather}
\frac{\partial f}{\partial t}+c_i\frac{\partial f}{\partial x_i}=\C (f,f),
\end{gather}
where
\begin{gather}
\C (f,f)=\frac{f_{\sc r}-f}{\tau}
\end{gather}
and
\begin{gather}
f_{\sc r}=f^{\sc (0)}\left(1+\phi_{\sc r}\right)=f^{\sc (0)}\left\{1+\tau\, (1-Z_{\sc 0})\left[\!\left(\frac{5}{3}-\gamma\right)\left(\beta C^2-\frac{3}{2}\right)
-(\gamma-1)\left(\frac{\E_{\sc s}}{kT}-\frac{\E}{kT}\right)\!\right]\frac{\partial v_r}{\partial x_r}\right.\nonumber\\[5mm]
+\,\left. \tau\left[\!\left(\beta C^2-\frac{5}{2}\right)+\left(\frac{\E_{\sc s}}{kT}-\frac{\E}{kT}\right)\!\right]\frac{C_i}{T}
\spc \left[\frac{\partial T}{\partial x_i}-\int_0^t \frac{\f\spc (t-t^\prime)}{\gamma}\frac{\partial T({\bf x},t^\prime)}{\partial x_i}\spc dt^\prime\right]\right\}.
\end{gather}
The multiplication of model equation (A.1) by an arbitary function $\psi=\psi ({\bf x},{\bf c},t,s)$ and subsequent integration of the resulting equation 
over all values of the molecular velocity ${\bf c}$ and internal degrees of freedom $s$ leads to the transfer equation
\begin{gather}
\frac{\partial }{\partial t}\int \psi\spc f\spc d{\bf c}\spc ds
+\frac{\partial }{\partial x_i}\int \psi\spc c_i\spc f\spc d{\bf c}\spc ds
-\int\left[\frac{\partial \psi}{\partial t}+c_i\frac{\partial \psi}{\partial x_i}
\right]f\spc d{\bf c}\spc ds=\int \psi\, \C (f,f)\spc d{\bf c}\spc ds.
\end{gather}
Note that the right-hand side of the above transfer equation vanishes if $\psi$ is a summational invariant.  
In the kinetic theory of gases, the specific entropy density $\sigma$ is usually defined as
\begin{gather}
\rho \sigma=-k\int f\ln (bf)\spc  d{\bf c}\spc ds,
\end{gather}
where $b$ is a constant which makes the argument of the logarithm function dimensionless. 
The balance equation for the entropy density $\rho \sigma$ can be derived from transfer equation (A.4) by 
taking $\psi=-k\ln (bf)$. Hence, we have  
\begin{gather}
\frac{\partial \rho \sigma}{\partial t}+\frac{\partial}{\partial x_i}\left(\rho\sigma v_i+\varphi_i\right)=\varsigma,
\end{gather}
where
\begin{gather}
\varphi_i=-k\int f\ln (bf)\spc C_i\spc d{\bf c}\spc ds
\end{gather}
is the entropy flux and 
\begin{gather}
\varsigma=-k\int \ln (bf)\spc \C (f,f)\spc d{\bf c}\spc ds=-\frac{k}{\tau}\int (f_{\sc r}-f)\ln (bf)\spc d{\bf c}\spc ds
\end{gather}
is the entropy production density. The model equation (A.1) satisfies an $H$-theorem if the entropy production density is a positive semi-definite 
quantity, i.e., $\varsigma \geq 0$. In order to proof this condition, we write entropy production density as 
\begin{gather}
\varsigma=-\frac{k}{\tau}\int f_{\sc r}\left(1-f/f_{\sc r}\right)\ln (f/f_{\sc r})\spc \spc d{\bf c}\spc ds
-\frac{k}{\tau}\int (f_{\sc r}-f)\ln (bf_{\sc r})\spc d{\bf c}\spc ds=\varsigma_{\sc 1}+\varsigma_{\sc 2}.
\end{gather}
It is easy to verify that $\varsigma_{\sc 1}$ is positive semi-definite due to the inequality $(1-x)\ln x \leq 0$ which is valid for all $x=f/f_{\sc r} > 0$ 
with the equality sign just if and only if $x=1$, i.e., when $f=f_{\sc r}$. The term $\varsigma_{\sc 2}$ may be rewritten as 
\begin{gather}
\varsigma_{\sc 2}=-\frac{k}{\tau}\int (f_{\sc r}-f)\ln (b f^{\sc (0)})\spc d{\bf c}\spc ds
-\frac{k}{\tau}\int (f_{\sc r}-f)\spc \phi_{\sc r}\spc d{\bf c}\spc ds.
\end{gather}
Since $\ln (b f^{\sc (0)})$ is a summational invariant, the first term on the right-hand side of (A.10) vanishes. Concerning the second term on the 
right-hand side of (A.10) we have
\begin{gather}
-\frac{k}{\tau}\int (f_{\sc r}-f)\spc \phi_{\sc r}\spc d{\bf c}\spc ds
=\frac{k}{\tau}\int (f-f^{\sc (0)})\, \phi_{\sc r}\spc d{\bf c}\spc ds
-\frac{k}{\tau}\int f^{\sc (0)}\phi_{\sc r}^2\spc d{\bf c}\spc ds.
\end{gather}
By using the conditions
\begin{gather}
\int (f-f^ {\sc (0)})\spc d{\bf c}\spc ds=0,
\end{gather}
\begin{gather}
\int C_i\spc (f-f^ {\sc (0)})\spc d{\bf c}\spc ds=0,
\end{gather}
\begin{gather}
\int \left[\left(\beta C^2-\frac{3}{2}\right)+\left(\frac{\E_{\sc s}}{kT}-\frac{\E}{kT}\right)\right](f-f^ {\sc (0)})\spc d{\bf c}\spc ds=0,
\end{gather}
\begin{gather}
k\int \left(\beta C^2-\frac{3}{2}\right)(f-f^ {\sc (0)})\spc d{\bf c}\spc ds=-\frac{3}{2}\spc 
\frac{\eta_{\sc 0}}{T}\spc \frac{\partial v_i}{\partial x_i},
\end{gather}
\begin{gather}
k\int \left[\left(\beta C^2-\frac{5}{2}\right)+\left(\frac{\E_{\sc s}}{kT}-\frac{\E}{kT}\right)\right]C_i\spc (f-f^ {\sc (0)})\spc d{\bf c}\spc ds=
-\int_0^t\frac{\lambda (t-t^\prime)}{T}\spc \frac{\partial T({\bf x},t^\prime)}{\partial x_i}\spc dt^\prime,
\end{gather}
we get
\begin{gather}
\frac{k}{\tau}\int (f-f^{\sc (0)})\, \phi_{\sc r}\spc d{\bf c}\spc ds=-(1-Z_{\sc 0})\spc \frac{\eta_{\sc 0}}{T}\spc \frac{\partial v_i}{\partial x_i}
\frac{\partial v_j}{\partial x_j}\nonumber\\[5mm]
-\,\frac{\gamma \mu_{\sc 0} c_{\sc v}}{\displaystyle T^{\spc 2}}
\left[\frac{\partial T}{\partial x_i}-\int_0^t \frac{\f\spc (t-t^\prime)}{\gamma}\frac{\partial T({\bf x},t^\prime)}{\partial x_i}\spc dt^\prime\right]
\int_0^t\frac{\f\spc (t-t^\prime)}{\gamma}\spc \frac{\partial T({\bf x},t^\prime)}{\partial x_i}\spc dt^\prime
\end{gather}
and
\begin{gather}
\frac{k}{\tau}\int f^{\sc (0)}\phi_{\sc r}^2\spc d{\bf c}\spc ds=\frac{(1-Z_{\sc 0})^2}{Z_{\sc 0}}\spc \frac{\eta_{\sc 0}}{T}\spc 
\frac{\partial v_i}{\partial x_i}\frac{\partial v_j}{\partial x_j}\nonumber\\[5mm]
+\, \frac{\gamma \mu_{\sc 0} c_{\sc v}}{\displaystyle T^{\spc 2}}
\left[\frac{\partial T}{\partial x_i}-\int_0^t \frac{\f\spc (t-t^\prime)}{\gamma}\frac{\partial T({\bf x},t^\prime)}{\partial x_i}\spc dt^\prime\right]
\left[\frac{\partial T}{\partial x_i}-\int_0^t \frac{\f\spc (t-t^\prime)}{\gamma}\frac{\partial T({\bf x},t^\prime)}{\partial x_i}\spc dt^\prime\right].
\end{gather}

\newpage

\noindent Thus, by combining (A.10), (A.11), (A.17) and (A.18) we have
\begin{gather}
\varsigma_{\sc 2}=-(1-Z_{\sc 0})\left(\frac{5}{3}-\gamma\right)\frac{\mu_{\sc 0}}{T}\spc \frac{\partial v_i}{\partial x_i}\frac{\partial v_j}{\partial x_j}
-\frac{\gamma \mu_{\sc 0} c_{\sc v}}{\displaystyle T^{\spc 2}}
\left[\frac{\partial T}{\partial x_i}
-\frac{1}{\gamma \mu_{\sc 0} c_{\sc v}}\int_0^{\spc t} \lambda\spc (t-t^\prime)\spc \frac{\partial T({\bf x},t^\prime)}{\partial x_i}\spc dt^\prime\right]
\frac{\partial T}{\partial x_i}.
\end{gather}
For polyatomic gases $\gamma < 5/3$ and $Z_{\sc 0} > 1$, so that the first term on the right-hand side of expression (A.19) 
is positive semi-definite. In our constitutive relation for the heat flux vector, the memory (or retardation) effect is associated with the 
coefficient of thermal conductivity so that the integral appearing in expression (A.19) can be approximated by perfoming the following Markoffian approximation 
\begin{gather}
\int_0^{\spc t} \lambda\spc (t-t^\prime)\spc \frac{\partial T({\bf x},t^\prime)}{\partial x_i}\spc dt^\prime
=\int_0^{\spc t} \lambda\spc (t^\prime)\spc \frac{\partial T({\bf x},t-t^\prime)}{\partial x_i}\spc dt^\prime
=\frac{\partial T({\bf x},t)}{\partial x_i}\int_0^\infty \lambda\spc (t^\prime)\spc dt^\prime.
\end{gather}
By inserting (A.20) into (A.19) we get
\begin{gather}
\varsigma_{\sc 2}=-(1-Z_{\sc 0})\left(\frac{5}{3}-\gamma\right)\frac{\mu_{\sc 0}}{T}\spc \frac{\partial v_i}{\partial x_i}\frac{\partial v_j}{\partial x_j}
+\frac{5}{4}\spc c_{\sc v} (\gamma-1)\spc \frac{\mu_{\sc 0}}{\displaystyle T^{\spc 2}}\spc
\frac{\partial T}{\partial x_i}\frac{\partial T}{\partial x_i},
\end{gather}
which is positive semi-definite. Thus we have proved that
\begin{gather}
\varsigma=\varsigma_{\sc 1}+\varsigma_{\sc 2} \geq 0,
\end{gather}
i.e., our generalized model equation fulfills the requirement of an $H$-theorem.

\newpage
\section{}

\noindent The non-zero elements of the matrices ${\bf M}^{\sc (0)}$ and ${\bf M}^{\sc (1)}$ read
\begin{gather}
\M_{\sc 11}^{\sc (0)}=W(z),\quad\quad \M_{\sc 12}^{\sc (0)}=2\spc A(z),\quad\quad \M_{\sc 13}^{\sc (0)}=B(z),
\end{gather}
\begin{gather}
\M_{\sc 21}^{\sc (0)}=\frac{\M_{\sc 12}^{\sc (0)}}{2},\quad\quad \M_{\sc 22}^{\sc (0)}=z\spc \M_{\sc 12}^{\sc (0)},\quad\quad \M_{\sc 23}^{\sc (0)}=
z\spc \M_{\sc 13}^{\sc (0)},
\end{gather}
\begin{gather}
\M_{\sc 31}^{\sc (0)}=(\gamma-1)\spc\M_{\sc 13}^{\sc (0)},\quad\quad \M_{\sc 32}^{\sc (0)}=2\spc (\gamma-1)\spc \M_{\sc 23}^{\sc (0)},
\quad\quad \M_{\sc 33}^{\sc (0)}=(\gamma-1)\spc D(z)+\frac{3}{2}\biggl(\frac{5}{3}-\gamma\biggr)\spc W(z),
\end{gather}
\begin{gather}
\M_{\sc 12}^{\sc (1)}=(1-Z_{\sc 0})\biggl(\frac{5}{3}-\gamma\biggr)B(z),\quad\quad \M_{\sc 13}^{\sc (1)}=
\biggl(1-\frac{\f\,(\omega)}{\gamma}\biggr)C(z),
\end{gather}
\begin{gather}
\M_{\sc 22}^{\sc (1)}=z\spc \M_{\sc 12}^{\sc (1)},\quad\quad \M_{\sc 23}^{\sc (1)}=
z\spc \M_{\sc 13}^{\sc (1)},
\end{gather}
\begin{gather}
\M_{\sc 32}^{\sc (1)}=(1-Z_{\sc 0})(\gamma-1)\biggl(\frac{5}{3}-\gamma\biggr)E(z),\quad\quad \M_{\sc 33}^{\sc (1)}=
\biggl(1-\frac{\f\, (\omega)}{\gamma}\biggr)\biggl[(\gamma-1)\spc F(z)+\frac{3}{2}\biggl(\frac{5}{3}-\gamma\biggr)\spc A(z)\biggr],
\end{gather}
with
\begin{gather}
\f\,(\omega)=\frac{\f_{\spc \sc 0}}{1-i\spc\omega\spc \tau_{\sc q}}=\f_{\spc \sc 0}\left(1-\frac{i}{\R}\spc \frac{\f_{\spc \sc 0}}{\gamma}\right)^{\!\! -1}
\end{gather}
being the frequency-dependent Eucken factor. Moreover, we have introduced the abbreviations
\begin{gather}
A(z)=z W(z)+1,\quad\quad B(z)=\left(z^2-\frac{1}{2}\right)W(z)+z,\quad\quad C(z)=z\left(z^2-\frac{3}{2}\right)W(z)+z^2-1,\\[5mm]
D(z)=\left(z^4-z^2+\frac{5}{4}\right)W(z)+z^3-\frac{z}{2},\quad\quad E(z)=\left(z^4-z^2-\frac{1}{4}\right)W(z)+z^3-\frac{z}{2},\\[5mm]
F(z)=z\left(z^4-2\spc z^2+\frac{7}{4}\right)W(z)+z^4-\frac{3}{2}\spc z^2+\frac{3}{2}.
\end{gather}
\end{appendices}

\newpage
\bibliographystyle{unsrtdin}

\end{document}